\documentclass[12pt]{article}

\usepackage{amsmath}
\usepackage{amssymb}
\usepackage[T1]{fontenc}
\usepackage[utf8x]{inputenc}
\usepackage[english]{babel}
\usepackage{placeins}
\usepackage{array}
\usepackage{titling}
\usepackage[colorlinks=true, allcolors=blue]{hyperref}
\usepackage{setspace}
\usepackage{enumitem}
\usepackage[rightcaption]{sidecap}
\usepackage[backend=biber, style=numeric, sorting=none,maxnames=10]{biblatex}
\usepackage{wrapfig}

\usepackage[T1]{fontenc}

\newcommand{\email}[1]{\href{mailto:#1}{\tt{\nolinkurl{#1}}}}
\newcommand{\orcid}[1]{ORCID: \href{https://orcid.org/#1}{\tt{\nolinkurl{#1}}}}

\usepackage[sfdefault,lf]{carlito}
\usepackage[parfill]{parskip}
\usepackage{fancyhdr}
\usepackage[affil-it]{authblk} 

\renewcommand{\abstractname}{} 

\usepackage[a4paper,top=3cm,bottom=2cm,left=3cm,right=3cm,marginparwidth=1.75cm]{geometry}
\usepackage{tocloft}
\usepackage{amsmath}
\usepackage{graphicx}
\usepackage{booktabs}

\usepackage{tcolorbox}

\usepackage{xr}
\usepackage[colorinlistoftodos]{todonotes}
\graphicspath{ {./figures/} }

\title{Network-driven discovery of repurposable drugs targeting hallmarks of aging}
\author[1,2]{Bnaya Gross}
\author[1,2]{Joseph Ehlert}
\author[3]{Vadim N. Gladyshev}
\author[3]{Joseph Loscalzo}
\author[1,2]{Albert-L\'aszl\'o Barab\'asi \thanks{Corresponding author: barabasi@gmail.com}}
\affil[1]{Network Science Institute, Northeastern University, Boston, MA
02115}
\affil[2]{Department of Physics, Northeastern University, Boston, MA 02115}
\affil[3]{Department of Medicine, Brigham and Women’s Hospital, Harvard Medical School, Boston, MA 02115}
\date{\footnotesize \today}
\begin{document}
\maketitle
\thispagestyle{fancy}

\vspace{-2cm}
\renewcommand{\abstractname}{}
\begin{abstract}
\textbf{Abstract:} Despite the thousands of genes implicated in age-related phenotypes, effective interventions for aging remain elusive, a lack of advance rooted in the multifactorial nature of longevity and the functional interconnectedness of the molecular components implicated in aging. Here, we introduce a network medicine framework that integrates 2,358 longevity-associated genes onto the human interactome to identify existing drugs that can modulate aging processes. We find that genes associated with each hallmark of aging form a connected subgraph, or hallmark module, a discovery enabling us to measure the proximity of 6,442 clinically approved or experimental compounds to each hallmark. We then introduce a transcription-based metric, $pAGE$, which evaluates whether the drug-induced expression shifts reinforce or counteract known age-related expression changes. By integrating network proximity and $pAGE$, we identify multiple drug repurposing candidate that not only target specific hallmarks but act to reverse their aging-associated transcriptional changes. Our findings are interpretable, revealing for each drug the molecular mechanisms through which it modulates the hallmark, offering an experimentally falsifiable framework to leverage genomic discoveries to accelerate drug repurposing for longevity.

\end{abstract}

\setstretch{2}

\newpage

\noindent \textbf{Introduction}

\noindent In the past decade, comprehensive genetic surveys \cite{de2009meta,newman2010meta,timmers2019genomics,timmers2020multivariate,zenin2019identification} and systematic animal experiments \cite{guarente2000genetic,mitchell2015animal} have implicated thousands of human genes in age-related phenotypes, offering unprecedented opportunities to dissect the molecular basis of longevity \cite{finch2001genetics,finch1997genetics,brooks2013genetics,alon2023systems}. Despite the sheer scale of these discoveries, we continue to lack treatments and interventions capable of modulating specific aging processes. This shortfall potentially stems from the multifactorial nature of aging and the functional and mechanistic interconnectedness of the molecular and genetic processes implicated in longevity, limiting the impact of any single intervention. 

\par

The multifactorial nature of longevity is often formalized through the “hallmarks of aging,” which demarcate multiple distinct age-related mechanisms, ranging from genomic instability to cellular senescence \cite{lopez2013hallmarks,lopez2023hallmarks,pun2022hallmarks}. Although each hallmark is intended to represent a distinct biological dimension, they are not fully distinct, and extensive cross-talk and synergy exist among them. Yet, current therapeutic interventions in clinical trials typically target only one or at most a few facets of aging. A comprehensive strategy for promoting longevity will likely require multiple interventions, each addressing different mechanisms (or hallmarks) of aging. Developing novel compounds to achieve this is a lengthy endeavor requiring a decade or more to reach clinical practice. An attractive alternative is to repurpose from the pool of over 6,000 clinically approved or experimental drugs, because some of these agents might effectively target specific aging processes \cite{barzilai2012critical,wang2022epigenetic,fong2024principal}. Indeed, most of these compounds have already undergone toxicity screening and possess well-characterized targets (and have known side effects), allowing for more rapid clinical development. The key challenge is to identify the compounds that can influence longevity—and specifically, the hallmark they target and the relevant molecular mechanism.

\par

To address this bottleneck in aging research, here we introduce a network medicine framework \cite{menche2015uncovering,barabasi2011network,han2008understanding,cohen2022complex} that allows us to integrate data on thousands of aging-associated genes, along with their network relationships, and the targets of all approved and experimental drugs, aiming to identify potential interventions that could affect longevity. Specifically, we begin with a library of 2,358 previously identified longevity-associated genes, which we map onto the human interactome. Strikingly, we find that specific hallmark-associated genes aggregate into a connected subgraph, forming distinct and statistically significant hallmark modules. The discovery of these modules is our first discovery, enabling us to apply established network medicine approaches for drug repurposing \cite{guney2016network,guthrie2023autocore,morselli2021network} and to evaluate the proximity of 6,442 clinically approved or experimental compounds to each hallmark module, thereby identifying candidates capable of perturbing specific aging phenotypes. 

Our second key advance is the introduction of a novel transcription-based metric, $pAGE$, that allows us to assess whether the expression changes induced by a drug reinforce or counteract known age-related expression changes, allowing us to distinguish beneficial interventions from those that may accelerate aging. This advance was missing in prior repurposing efforts \cite{guney2016network,guthrie2023autocore,morselli2021network}, that only predicted the drug's impact on a given phenotype but not the directionality and the magnitude of its action. Ultimately, we find multiple drugs that successfully reverse the expression changes observed during aging in specific hallmarks, representing promising repurposing candidates. Our predictions are interpretable, revealing the precise molecular mechanisms by which each drug-repurposing candidate modulates the specific hallmark of aging, thereby providing experimentally falsifiable hypotheses. Together, our study offers a principled, integrative route to leverage the vast body of aging-related knowledge to identify drugs that can address the multifactorial nature of aging.

\par
\noindent
\underline{\textbf{Results:}}

\noindent \textbf{The genetic roots and interconnectivity of the hallmarks of aging}

\noindent We began by querying the OpenGenes Database \cite{rafikova2023open}, which curates gene-level annotations that link 2,358 genes to longevity, age-associated diseases, or pathways implicated in aging, also offering a confidence level for each association (Fig.~\ref{fig:Aging genes}\textbf{a} and Supplementary Section SI.I), ranging from 1 (highest) to 5 (lowest). 
From this resource, we identify the genes that are explicitly linked to at least one of the 11 hallmarks of aging \cite{lopez2013hallmarks,lopez2023hallmarks,pun2022hallmarks} (Fig.~\ref{fig:Aging genes}\textbf{b}).
Among the 2,358 longevity-associated genes, 1,250 can be associated to specific hallmarks of aging: 860 are exclusive to a single hallmark, and 390 span multiple hallmarks (Fig.~\ref{fig:Aging genes}\textbf{c}). The remaining 1,108 genes, while linked to aging, could not be associated with a specific hallmark based on the current knowledge of their biological function. We can, however, rely on the topology of the human interactome to link these genes to specific hallmarks (see Supplementary Section SI.III). The 390 multi-hallmark genes support our hypothesis that, at the molecular level, the hallmarks of aging are not independent entities. Further support is offered by pairwise comparisons using the Jaccard index \cite{jaccard1901etude}, revealing statistically significant gene overlap among 47 of the 55 hallmark pairs (Fig.~S7 and Supplementary Section SI.IV).
\par
\noindent Although the OpenGenes database offers evidence extracted from observational and experimental studies that link each gene to its assigned hallmark of aging, we wished to confirm whether the collective set of 1,250 hallmark-associated genes is broadly relevant to longevity. To this end, we performed additional validation (Supplementary Section SI.II and Methods), finding that the 1,250 gene set shows significant enrichment in (i) age-related KEGG pathways \cite{kanehisa2000kegg} (Fig.~S11), (ii) genes implicated in aging by seven large-scale aging studies \cite{timmers2019genomics,timmers2020multivariate,zenin2019identification,jia2018analysis,peters2015transcriptional,aging2021aging,sebastiani2021protein} (Table~S1), (iii) five aging-related diseases (Table~S2), (iv) eight distinct cancer types whose incidence increasing significantly with age (Fig.~S4), and (v) genes involved in DNA repair or progeroid syndromes \cite{freitas2011data,kipling2004can,burla2018genomic} (Fig.~S5). These enrichments lend additional support to the aging relevance of this gene set, serving as the foundation for our subsequent work.

\par \noindent \textbf{The hallmark modules of aging}

\noindent To capture the network-level organization of aging, we mapped the 1,250 hallmark-associated genes onto the human interactome—a comprehensive catalog of 524,156 experimentally validated binding interactions among 18,223 proteins (see Methods). For many diseases and phenotypes, the genes associated with the disease are known to coalesce in the interactome to form a \emph{disease module}, formally defined as the largest connected component (LCC) formed by the disease genes \cite{menche2015uncovering}. While disease modules were validated for multiple isolated traits \cite{sharma2015disease,do2021network,spector2025transformers,kersting2024nextflow}, aging comprises multiple hallmarks that may each behave as distinct, yet interrelated, phenotypes. It is thus unclear whether the different hallmark genes form independent modules, and if these modules reside in the same network neighborhood (Fig.~\ref{fig:hallmarks_relations}\textbf{a}).

\par 

To answer these questions we examined each hallmark separately, finding that in nine of the eleven hallmarks, associated genes cluster into a statistically significant LCC (z-score $> 1.96$, Fig.~\ref{fig:hallmarks_relations}\textbf{b}). The remaining two hallmarks—Loss of proteostasis (z-score $= 1.74$) and Epigenetic alterations (z-score $= 1.67$)—also show a marginal significance, indicating that their gene sets are characterized by non-random connectivity. In other words, the hallmark-associated genes reside in narrowly defined network neighborhoods, each representing a distinct and identifiable \emph{hallmark module} within the global interactome (Fig.~\ref{fig:longevity_module}\textbf{a–k}). This finding represents our first key discovery, establishing that for each hallmark, the hallmark genes form well-defined and statistically significant network modules.
\par
\noindent We next examined whether these hallmark modules overlap, aiming to reveal functional relationships among them. To do so, we used two complementary measures: \emph{separation} \cite{menche2015uncovering} and \emph{proximity} \cite{guney2016network} (Fig.~\ref{fig:hallmarks_relations}\textbf{a}, see also Methods and Supplementary Section SI.V), finding that the hallmarks of aging are located in the same neighborhood of the interactome, together forming a broader “\emph{longevity module}” (Fig.~\ref{fig:longevity_module}\textbf{l} and Supplementary Section SI.VI). Next, we leverage the existence of the individual hallmark modules to identify drug-repurposing candidates that target specific hallmarks.

\par \noindent \textbf{Network-based identification of hallmark-specific drug-repurposing candidates}

\noindent The existence of distinct hallmark modules offers the opportunity to apply network-based drug-repurposing methods, originally designed for single-disease modules, to the more complex and multifactorial context of aging \cite{guney2016network,guthrie2023autocore,morselli2021network}. To this end, we compiled 6,442 approved or clinically tested compounds from DrugBank \cite{wishart2006drugbank}. Our approach rests on the premise that drugs whose targets lie in the network proximity to a disease (or hallmark) module are poised to perturb that disease (or hallmark) with potential therapeutic outcome, a hypothesis that has been experimentally supported across multiple diseases—from asthma to heart disease, and has been experimentally validated for 6,710 drugs, successfully predicting their potential role in treating COVID-19 infection \cite{morselli2021network,sharma2015disease,do2021network,patten2022identification}.

\par We begin by measuring each drug’s network proximity to every hallmark module for five sets of hallmark genes, stratified by confidence level (Fig.~\ref{fig:Aging genes}\textbf{a}). For each hallmark, we then ranked drugs by the significance of their proximity (z-score$<-1.96$). For example, we find 26 compounds that displayed statistically significant proximity to the \emph{Cell senescence} hallmark at every confidence level (Table~\ref{table:top_drug_repurposing_each_hallmark}). The top-ranked candidate in this list, pimasertib, is a MEK1/2 inhibitor known to induce apoptosis and senescence \cite{hata2017synergistic}, aligning with its predicted effect. Another high-ranking compound, selisistat, a SIRT1 inhibitor, also promotes senescence \cite{fu2021srt1720}. These examples demonstrate that proximity successfully detects compounds previously implicated in senescence. Yet they also show that proximity alone does not imply a beneficial (anti-senescence) effect.
\par
The network-based approach relies on undirected protein interactions, which can successfully establish a compound’s ability to perturb a module, but carries no information on whether the perturbation is beneficial or detrimental. To address this limitation, here we introduce a metric called \emph{Pro-Age} ($pAGE$, see Box~1), which quantifies whether drug-induced changes in gene expression reinforce or counter documented age-related expression shifts. Specifically, if a drug up-regulates a gene that is known to be up-regulated with age, one might anticipate a potentially adverse effect on longevity, whereas down-regulating the same gene may be advantageous (Box.~1).
\par
In summary, we propose \textit{SHARP} (Systematic Hallmark-based Aging Repurposing Pipeline), which consists of two steps: (1) We rely on network proximity to identify compounds whose targets lie in the proximity of specific hallmark-related subgraphs, thereby filtering out thousands of compounds that target regions not associated with longevity. (2) We then measure each proximal compound’s $pAGE$ parameter to determine whether it reinforces or counteracts aging-related transcriptional changes, enabling us to distinguish potential "pro-longevity" drugs from “age-accelerating” agents.

\par \noindent \textbf{Validation of the SHARP Repurposing Pipeline}

To evaluate the validity and limitations of our drug-repurposing pipeline, SHARP, we first assess its ability to predict drugs whose relevance to aging has been supported by experimental and clinical evidence. Specifically, we examine whether SHARP can identify the 17 drugs currently in clinical trials for longevity \cite{guarente2023human} and the 11 compounds that have been experimentally shown to extend lifespan in mice \cite{miller2007aging}.

\par \underline{\textit{Validation 1: Drugs under clinical trials for humans.}} \noindent We first tested our drug-repurposing pipeline against a curated list of 17 compounds currently under clinical trial for healthy longevity \cite{guarente2023human}, a list that included well-known candidates such as metformin and sirolimus (rapamycin) (Table~\ref{table:clinical_ITP_drug_repurposing} and Table~S4). Of these 17, we find that 11 display statistically significant proximity to at least one hallmark (z-score$<-1.96$). For instance, Aspirin is predicted to influence six hallmarks, and dasatinib is predicted to affect five, whereas sirolimus (rapamycin) affects only one hallmark, specifically the \emph{Intercellular communication} hallmark module. The six drugs under clinical trials not captured by our pipeline had targets located relatively far from hallmark modules (proximity$>1.6$, Fig.~S14). Even so, three of the six compounds acarbose, metformin, and quercetin, exhibited marginally significant proximity (z-score $<-1.645$), illustrating that partial alignment with a hallmark can still be detected for these compounds. 
\par
Finally, we measure the $pAGE$ parameter for nine of the seventeen drugs in clinical trials for aging or longevity for which CMap data are available (see Methods), finding that all nine displayed positive $pAGE$ for at least three hallmarks (see Table~\ref{table:clinical_ITP_drug_repurposing} and Table~S4), indicating that they alleviate age-related expression changes. 

\par \underline{\textit{Validation 2: Drugs extending lifespan in mice (ITP).}} \noindent We next examined eleven drugs experimentally tested by the Intervention Testing Program (ITP) \cite{miller2007aging}, finding that they prolong lifespan in mice (Table~\ref{table:clinical_ITP_drug_repurposing} and Table~S3). Three of these (sirolimus, acarbose, and metformin) overlap with the seventeen compounds in the human clinical trials set above. Of the eleven, six displayed significant proximity (z-score$<-1.96$), and four had marginal proximity (z-score$<-1.645$) to at least one hallmark.  
\par
Finally, we measure the $pAGE$ parameter for eight of the eleven ITP-confirmed lifespan-extending drugs \cite{miller2007aging} with CMap data, finding that all eight have positive $pAGE$ for at least three hallmarks (Table~\ref{table:clinical_ITP_drug_repurposing} and Table~S3).
\par
Taken together, we find that SHARP captured 82.4\% of clinically tested compounds and 90.9\% of mouse lifespan-extending compounds, if we count both strong and marginal hits, and that the $pAGE$ measure indicates that each drug under clinical trials or with impact on mice lifespan induce expression changes that act to restore the age-induced changes in the respective hallmark module. These findings not only confirm the predictive power of our approach, but offer confidence in the novel predictions of drug repurposing candidates that we discuss below.
\par 

\noindent \textbf{Identifying hallmark-targeted drug repurposing candidates}

Encouraged by the positive validation above, we next apply SHARP to identify drug--repurposing opportunities for each hallmark. Specifically, we identify all approved and experimental compounds with significant proximity to specific aging modules, as identified by our network analysis, and positive $pAGE$ value, if there is available CMap data for the compound. This process allows us to identify \textit{pro-longevity compounds} that can successfully perturb a hallmark module, inducing pro-longevity expression changes. We also identify \textit{age-accelerating compounds} that can also perturb a hallmark, but the induced expression changes are more consistent with pro-aging effects.
\par
The network-based repurposing pipeline has allowed us to identify a total of 370 drugs (Table~\ref{table:top_drug_repurposing_each_hallmark}), each showing statistically significant proximity to one or more hallmarks of aging, hence potentially capable of modulating longevity. Of these, 60 drugs have CMap expression profiles, allowing us to calculate their $pAGE$ parameter. Among these 60 drugs, 14 display positive $pAGE$, indicating that they represent pro-longevity drugs and another 14 display negative $pAGE$, representing potential age-accelerating compounds. The remaining 32 drugs show inconsistent $pAGE$ values across the 5 confidence levels, hence we need further data to evaluate their impact on longevity. Below, we summarize the identified candidate drugs in the context of each hallmark.

\par \underline{\textit{Exhaustion of stem cells:}} 
\noindent We identified 113 drugs with significant proximity (z-score $<-1.96$) to this hallmark at confidence levels 3–5, 19 of which have CMap data. Four of 19 exhibit positive $pAGE$ across all tested levels (guanadrel, nisoxetine, amineptine, and amlexanox), and 5 are predicted to be age-accelerating compounds (protriptyline, iobenguane, enalaprilat, doramapimod, and benzatropine).

\par \underline{\textit{Altered intercellular communication:}} \noindent 61 drugs are significantly proximal to this hallmark across all five confidence levels, of which 25 have CMap data. Seven exhibit positive $pAGE$ across the board (oxymetazoline, metaraminol, terazosin, tamsulosin, tetryzoline, cirazoline, and synephrine) and seven are age-accelerating compounds (niguldipine
sertindole, doxazosin, naphazoline, linsitinib, bms-754807, and dequalinium).

\par \underline{\textit{Epigenetic alterations:}} \noindent Of 52 drugs with significant proximity across all five confidence levels, only five have CMap data. Of these, clinofibrate has positive $pAGE$ across all levels, while pilaralisib is predicted to be an age-accelerating compound.

\par \underline{\textit{Mitochondrial dysfunction:}} \noindent 21 drugs exhibit significant proximity across all five confidence levels, but none have CMap data. If we also consider drugs that show proximity for four out of five confidence levels, we find three drugs with CMap data (navitoclax, alsterpaullone, and pyrazolanthrone). Among them, pyrazolanthrone has positive $pAGE$ across all levels.

\par \underline{\textit{Loss of proteostasis:}} \noindent 6 drugs have significant proximity at all five confidence levels but we do not have CMap data for any of them. Considering those display proximity at four out of five confidence levels, we find minocycline with CMap data, however, its $pAGE$ values are inconsistent at the different confidence levels.

\par \underline{\textit{Changes in the extracellular matrix structure:}} 
\noindent As no genes from the OpenGenes database have confidence levels 1 or 2 for this hallmark, we examined levels 3–5, identifying 25 significantly proximal drugs. Among these, two drugs have CMap data: marimastat with a positive $pAGE$ and captopril with a negative $pAGE$. 

\par \underline{\textit{Deregulated nutrient sensing:}} 
\noindent 52 drugs reach significance (z-score $<-1.96$) across all five confidence levels, four of which have CMap data. While none show positive $pAGE$ at all five levels, three drugs exhibit positive $pAGE$ in four significance levels (bms-754807, pilaralisib, and linsitinib).

\par \underline{\textit{Genomic instability:}} \noindent 4 drugs achieve significant proximity at all five confidence levels, yet none have CMap data. Adding those significant at four levels yields three drugs with CMap data (gsk-1059615, paricalcitol, and pimecrolimus), of which gsk-1059615 and pimecrolimus are age-accelerating compounds exhibiting negative $pAGE$ across all levels.

\par \underline{\textit{Cell senescence:}} \noindent Twenty-seven drugs are significant at all five levels; three of them-biotin, linsitinib and bms-754807 have CMap data, but show inconsistent $pAGE$ values.

\par \underline{\textit{Disabled macroautophagy:}} \noindent With no confidence level 1 genes available, we analyzed levels 2–5, identifying seven drugs that reach significance for all four. Two, monobenzone and imexon, have CMap data and imexon also has a positive $pAGE$ value.

\par \underline{\textit{Telomere attrition:}} \noindent Two drugs prove significant across all five levels, though neither has CMap data.

\par To summarize, we find 370 drugs that exhibit significant proximity to at least one hallmark of aging. Of the 370 drugs, 60 have CMap data, enabling us to compute their $pAGE$; of these 14 have a positive $pAGE$ across all five confidence levels, making them prime candidates for experimental testing in animal models. We also find additional 14 compounds with negative $pAGE$, indicative of potential age-accelerating effects. 
\par
We wish to emphasize that 310 of our drug-repurposing predictions currently lack CMap data. One can rely, therefore, on expression profiling to determine their $pAGE$ value, and assess their directionality. Extrapolating from our previous data, we anticipate that 23.3\% (or about 72 drugs) of these candidates may benefit longevity (see Methods).

\noindent

\noindent \textbf{Proximity and $pAGE$ predict therapeutic effects}

The integrated network-based pipeline, augmented by the $pAGE$ metric, is not only capable of identifying promising drug-repurposing candidates, but can also yield falsifiable predictions pertaining to the drug's mechanism of action. We demonstrate this on oxymetazoline (see other candidates in Supplementary Section SI.XI), a repurposing candidate that according to our pipeline impacts the Altered Intercellular Communication hallmark (Table.~\ref{table:top_drug_repurposing_each_hallmark}). 
Oxymetazoline is an adrenergic $\alpha_1$- and $\alpha_2$-agonist and a direct-acting sympathomimetic drug and is available in various formulations with a wide variety of clinical implications, including nasal congestion, allergic reactions of the eye, and facial erythema associated with rosacea \cite{patel2017oxymetazoline}. Oxymetazoline targets the proteins ADRA1A, ADRA1B, and ADRA1D, members of the $\alpha_1$-adrenergic receptor protein group, and the proteins HTR1A, HTR1B, and HTR1D, members of the serotonin receptor protein group (Fig.~\ref{fig:MOA}\textbf{a}). Of these, ADRA1A is a hallmark gene (confidence level 1), as mice expressing a constitutively active mutant ADRA1A lived significantly longer \cite{doze2011long}. While the potential impact of oxymetazoline on longevity is unknown, perturbing the activity of ADRA1A has the potential to extend lifespan by altering molecular mechanisms related to insulin signaling, the AMPK and TOR pathways, and chronic inflammation \cite{lagunas2022g}.
\par
To understand how the perturbation induced by oxymetazoline propagates through the hallmark module, we examined the gene perturbation signature in the vicinity of its targets: the $\alpha_1$-adrenergic receptor protein group ADRA1A, ADRA1B, and ADRA1D, and the serotonin receptor protein group HTR1A, HTR1B, and HTR1D. The $\alpha_1$-adrenergic receptor protein group directly connects to the module by the ALB and NR3C1 genes, while the serotonin receptor protein group directly connects to the module by the TGFB1 gene. The ALB, NR3C1 and TGFB1 proteins are direct interacting partners of oxymetazoline's targets in the hallmark module. Yet, the drug-induced perturbation signatures of the hallmark proteins ALB, NR3C1, and TGFB1 are weak (z-score $=0.39$ for ALB, z-score $=-0.18$ for NR3C1, and z-score $=0.36$ for TGFB1). This suggests that the drug induced perturbation propagates through the ACKR3 protein, a non-hallmark protein, which has some of the highest perturbation scores of the targets' neighbors (z-score $=-0.66$, Fig.~\ref{fig:MOA}\textbf{b}). The ACKR3 protein interacts with the hallmark proteins NFKB1 (z-score $=0.77$), TP53 (z-score $=1.14$), and AKT1 (z-score $=-0.79$), each displaying significant perturbation signature. While the expression patterns of the $\alpha_1$-adrenergic receptor proteins do not change with age (Fig.~\ref{fig:MOA}\textbf{c}), we predict that targeting them with oxymetazoline leads to a perturbation signature that affects the expression patterns of multiple genes in the Altered intercellular communication hallmark module (Fig.~\ref{fig:MOA}\textbf{b}) resulting in a statistically significant $pAGE = 0.46$ (z-score $= 2.35$, see Supplementary Section SI.VIII). Specifically, the expression pattern of genes CCL5, NFE2L2, AGER, RELA, CEBPB, C3, MIF, NFKBIA, HDAC4, PTGS2, HLA-DRB1 that are involved in the aging mechanism of sterile inflammation, are perturbed by oxymetazoline in the direction that corrects the aging-induced expression changes (Fig.~\ref{fig:MOA}\textbf{b,c}), increasing the $pAGE$ value (Fig.~\ref{fig:MOA}\textbf{d}). Similarly, the expression of CCL5, TP63, NFE2L2, FOXO3, CDKN2B, IGF1R, and SIRT1 genes involved in the aging mechanism of intercellular communication impairment, are also perturbed by oxymetazoline, opposing their aging-induced expression changes and further increasing the $pAGE$ value (Fig.~\ref{fig:MOA}\textbf{b-d}). 
\par
In summary, the integration of the network module (Fig.~\ref{fig:MOA}\textbf{a}), the drug's perturbation profile (Fig.~\ref{fig:MOA}\textbf{b}), and age-associated expression changes (Fig.~\ref{fig:MOA}\textbf{c}) unveils the molecular mechanism by which a repurposable drug is expected to modulate a hallmark module. The predicted mechanism can then be validated in cell-based assays \cite{sharma2015disease,do2021network} and in appropriate animal models. We wish to emphasize, however, that our focus on oxymetazoline is intended to serve as an illustrative example; the integration of the $pAGE$ metric with the network structure of the respective hallmark (Fig.~\ref{fig:longevity_module}) enables us to generate similarly detailed mechanistic predictions for each drug in Table.~\ref{table:top_drug_repurposing_each_hallmark}, that we predict to modulate longevity (see other candidates in Supplementary Section SI.XI).

\noindent

\noindent \textbf{Discussion}\\
\noindent An important paradigm in aging research is the distinction between the \emph{why} of aging, represented by causal factors \cite{gensler1981dna,schumacher2008age,pal2016epigenetics} and the \emph{how} of aging, as encapsulated by “aging hallmarks” \cite{gems2021hoverfly}. The longevity module introduced here reveals that from a network perspective, both the causal factors and the "hallmarks" of aging are located in the same network neighborhood. This implies that therapeutic interventions designed to target either the drivers of aging or its hallmark processes must ultimately focus on the same well-localized neighborhood of the sub-cellular network, defined by the longevity module (Fig.~\ref{fig:longevity_module}). Consequently, from a network perspective, the traditional distinction between targeting “theories” or “hallmarks” may be less critical, given the realization that both involve the same network neighborhood.
\par 
Ultimately, our findings underscore the potential of leveraging the extensive hallmark-associated genetic evidence to identify drug-repurposing candidates for healthy longevity. Although the evidence presented here is primarily computational, it is supported by genetics and expression-based evidence, offering a principled basis for subsequent \textit{in vitro} and \textit{in vivo} validations, culminating in animal studies and eventually clinical trials.
\par 
We also introduce a key methodological advance, the $pAGE$ metric, that helps us gauge whether a drug reinforces or counteracts aging-related transcriptional changes. Consequently, our pipeline uncovers both compounds that can act as pro-longevity compounds, as well as compounds that likely serve as “age-accelerating” agents. These are also valuable for validating the genetic nexus of aging, illuminating further molecular targets, and identifying previously unknown side-effects of existing drugs, helping clinicians to avoid unintended adverse effects on lifespan. 
\par
Our focus here has been on compounds that can modulate individual hallmarks of aging. However, given that aging is a multifactorial phenomenon, it is unlikely that a single drug can successfully perturb and alter all its signatures. Instead, multiple interventions—potentially involving combination therapies—will be required. Notably, our approach also identifies drugs that perturb several hallmarks, thereby offering a framework for multi-target strategies (Tables~S5 and S6). Future work should also explore rational combinations of drugs, resulting in therapeutic cocktails that can target multiple hallmarks simultaneously. Network-based models that factor in drug–drug interactions, synergy, and toxicity could inform such combination therapies \cite{cheng2019network}.  
\par 
Our approach has several limitations. First, aging processes vary substantially across tissues and cell types \cite{goeminne2024plasma}, necessitating the integration of tissue-specific or single-cell data to improve hallmark definition and refine the tissue-level $pAGE$ metric. Such tissue dependence can be systematically integrated into our repurposing pipeline by leveraging the GTEx (Genotype-Tissue Expression) database \cite{lonsdale2013genotype}. This enables us to filter out proteins not expressed in specific cell types and re-assess a drug’s potential as a perturbant within a more biologically relevant network context \cite{goeminne2024plasma}.
\par
Regarding our second key advance, the $pAGE$ parameter, we note that it does not yet account for dosage, non-linear responses, or the possibility that a single compound may confer beneficial effects in one tissue and detrimental effects in another. Future work could help improve its predictive value by implementing these features.
\par
Capturing the progression of hallmark modules at different chronological ages in humans or model organisms could also reveal critical windows when interventions are most effective, guiding stage-specific therapeutics for healthy aging.  Finally, combining patient stratification (e.g., genetic background, lifestyle factors) with network-derived hallmark signatures might help tailor interventions to individual aging trajectories, moving the field toward personalized anti-aging strategies and opening the doors for \textit{Precision Geroscience}.
\newpage
\par
\underline{\textbf{Online Methods:}}\\
\noindent
\textbf{The human interactome}\\
\noindent
The human PPI is assembled using experimentally validated protein interactions including (i) binary interactions, derived from high-throughput yeast two-hybrid experiments, three-dimensional protein structures; (ii) interactions identified by affinity purification followed by mass spectrometry; (iii) kinase substrate interactions; (iv) signaling interactions; and (v) regulatory interactions. The final PPI used in our study consists of 18,223 proteins connected by 524,156 binding interactions.
\par
\noindent
\textbf{LCC statistical significance}\\
\noindent
A set of genes $A$ with a degree distribution $P_A(k)$ forms the largest connected component $LCC(A) \in A$ in the interactome. The degree distribution of the interactome $P_G(k)$ is sampled using log-binning with a bin size of 100 nodes. The statistical significance of $LCC(A)$ is measured by random sampling 1000 gene sets in the interactome with size $|A|$ and degree distribution $P_A(k)$ and measuring the expected distribution of the LCC sizes resulting in a z-score for the $LCC(A)$ statistical significance.  
\par
\noindent
\textbf{Network-based separation}\\
\noindent
The network-based separation $S(A, B)$ between pairs of genes set $A$ and $B$ is calculated using the separation measurement introduced in Ref.~\cite{menche2015uncovering} 
\begin{equation}
    S(A,B) = \langle d_{AB} \rangle - \frac{\langle d_{AA} \rangle + \langle d_{BB} \rangle}{2}
\end{equation}
where $\langle d_{AB} \rangle$ is the average shortest path between proteins in different gene sets and $\langle d_{AA} \rangle$ and $\langle d_{BB} \rangle$ are the average shortest path between proteins within the same gene set.
\par
\noindent
\textbf{Network-based proximity}\\
\noindent
The network-based proximity $P(S, T)$ between pairs of genes set $S$ and $T$ (pair of hallmarks of aging, or a hallmark of aging a drug targets) is calculated using the proximity measurement introduced in Ref.~\cite{guney2016network}
\begin{equation}
    P(S,T) = \frac{1}{||T||} \sum_{t \in T} \min_{s \in S} d(s,t)
\end{equation}
where $d(s,t)$ is the  shortest  path  length between nodes $s$ and $t$ in  the network. The statistical significance of the proximity is obtained by comparing $P(S,T)$ to the distribution of proximity values of 1000 random selections of two sets of genes with size and degree distribution similar to $S$ and $T$.
\par
\noindent
\textbf{Analysis of gene expression and perturbation parameters}\\
\noindent
The perturbation signatures of genes in the hallmark modules are retrieved from the Connectivity Map (CMap) database (\href{https://clue.io/}{https://clue.io/}) for the MCF7 cell line after treatment with all drugs in the CMap database. These signatures reflect the perturbation of the gene expression profile in the hallmark modules caused by the treatment with that particular drug relative to a reference population, which comprises all other treatments in the same experimental plate \cite{subramanian2017next}. For drugs having more than one experimental instance (such as time of exposure, cell line, and dose), the one with the highest distil\_cc\_q75 value is selected (75th quantile of pairwise  Spearman correlations in landmark genes, \href{https://clue.io/connectopedia/glossary}{https://clue.io/connectopedia/glossary}). 
\par
\noindent
\textbf{Gene disease associations}\\
\noindent
By surveying over 120 databases with Gene-Disease-Associations (GDA) we selected those that i) were not compiled from other data sources, and ii) provided at least one kind of evidence type classified as Strong (functional evidence using an experimental essay); Weak (GWAS evidence but no experimental validation); Inferred (relying on bioinformatics or SNPs from imputation in GWAS); not compatible [(I)ncRNA, miRNA and other transcripts with or without experimental validation]. For each database we kept the disease name, gene converted to HGNC names (HUGO Gene Nomenclature Committee), and evidence level. Finally, we combined the following data sources: GWAS from ClinGen, ClinVar, CTD, Disease Enhancer, DisGeNET, GWAS Catalog, HMDD45, lncBook, LncRNA disease, LOVD, Monarch, OMIM, Orphanet, PheGenI, and PsyGeNet. All types of association were used in this study.
\par
\noindent
\underline{\textbf{Data availability:}}\\
\noindent
All data supporting the findings of this study are available at (https://github.com/BnayaGross/Longevity-module) and within the paper and its Supplementary
Information files.
\par
\noindent
\underline{\textbf{Code availability:}}\\
\noindent
The code used to produce the results and figures in the study is published on GitHub (https://github.com/BnayaGross/Longevity-module).
\par
\noindent
\underline{\textbf{Acknowledgements:}}\\
\noindent
A.L.B is supported by the European Union’s Horizon 2020 research and innovation program under grant agreement No. 810115 – DYNASNET. J.L. is supported in part by NIH grants R01 HL155107, R01 HL166137, and U01 HG007691; American Heart Association grants 957729 and AHA24MERIT 1185447; EU Horizon 2021 grant 101057619 (RePO4EU) to J.L. B.G. acknowledges the support of the Fulbright Postdoctoral Fellowship Program. We thank F. Nasirian for helpful discussions.
\par
\noindent
\underline{\textbf{Author contributions:}}\\
\noindent
B.G. performed data query and integration, statistical modeling, network analysis, programming, conceptual design, and writing the manuscript. J.E. contributed to network analysis, and statistical modeling. V.-N.G contributed to interpreting the results. J.L. contributed to interpreting the results and writing the manuscript. A.-L.B. contributed to the conceptual design of the study and writing the manuscript.

\par
\noindent
\underline{\textbf{Competing interests:}}\\
\noindent
A.L.B and J.L. are founders of Scipher Medicine, Inc., which applies network medicine strategies to biomarker development and personalized drug selection; B.G., J.E., and V.N.G. declare no conflicts.

\FloatBarrier

\newpage

\begin{tcolorbox}[colback=blue!10!white, colframe=blue!50!black, title=Box 1: Pro-Age ($pAGE$) Metric]
\footnotesize The network medicine framework relies on undirected protein interactions, which capture a drug's ability to perturb disease module, but lack information about the direction of the induced change, or whether a drug-induced perturbation is beneficial or detrimental for the studied phenotype. To overcome this limitation, we introduce the $pAGE$ metric, which quantifies whether a drug's impact on gene expression reinforces or counteracts documented age-related shifts. The metric is defined in three steps: (i) We define the \emph{longevity vector} $\Theta = \{\sigma'_g\}_{g \in \Lambda}$, that encodes the age-induced expression change of 2,025 genes, where $\sigma'_g = +1$ for 995 genes that are up-regulated with age (blue, left panel) according to the OpenGenes database and $\sigma'_g = -1$ for the 1,030 genes that are down-regulated with age (red, left). (ii) We define a drug’s perturbation signature $\Gamma = \{\sigma_g\}_{g \in \Lambda}$, for a set of genes $\Lambda$, capturing each gene's changes in expression following exposure to the drug according to the Connectivity Map (CMap) \cite{subramanian2017next}. Specifically, $\sigma_g \in \{-1, 0, +1\}$ in $\Gamma$ indicates whether drug exposure decreases (red, right panel), leaves unchanged (white), or increases (blue) expression of gene $g$.
    \begin{wrapfigure}{l}{0.58\textwidth} \vspace{-0.5cm}
        \centering
        \includegraphics[width=0.58\textwidth]
        {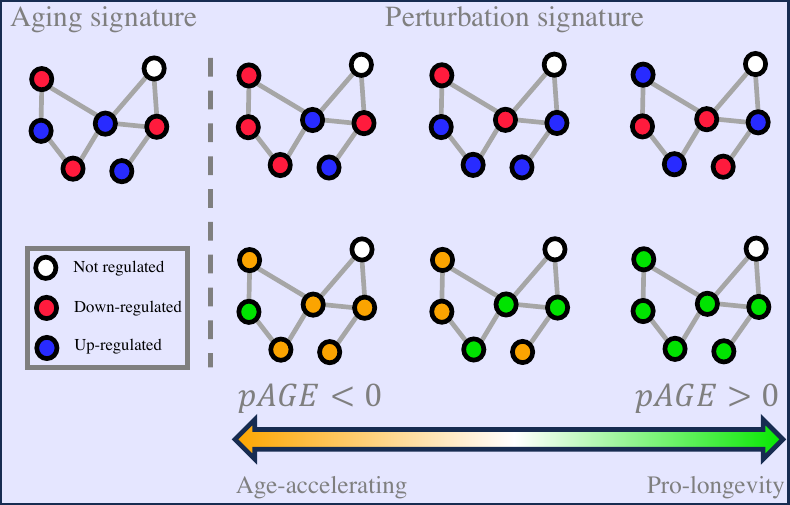}     
    \end{wrapfigure}
(iii) Finally, to asses the alignment between a drug’s perturbation profile  $\Gamma$ and the longevity vector $\Theta$, we introduce the $pAGE$ measure, defined as
\begin{equation} 
pAGE = -\frac{1}{|\Lambda_\emptyset|} \sum_{i \in \Lambda} \sigma_i \sigma'_i, 
\label{eq:pAGE} 
\end{equation} 
where $\Lambda_\emptyset = \{i \mid \sigma_i \sigma'_i \neq 0\}$ is a normalization factor that constrains $pAGE$ to vary between $-1$ and $+1$ (bottom panel). Whenever a drug's induced expression changes matches an age-related directional shift ($\sigma_i$ and $\sigma'_i$ share the same sign), the term $\sigma_i \sigma'_i$ is positive, pushing $pAGE$ toward negative values (orange). Conversely, if the drug induces down-regulation of a gene that is typically up-regulated with age, it turns $\sigma_i \sigma'_i$ negative, increasing $pAGE$ (green). Thus, $pAGE > 0$ indicates that a drug attenuates age-related expression changes (pro-longevity compound), whereas $pAGE < 0$ implies that it may exacerbate them (age-accelerating compound). The statistical significance of the $pAGE$ value is evaluated by comparing it to a control group of random drug signatures (see Supplementary Section SI.VIII).
\end{tcolorbox}


\begin{SCfigure}[1][h]
\caption{\textbf{Aging-associated genes. a,} The OpenGenes database contains 2,358 aging-associated genes, each gene being assigned a confidence level ranging from 1 (Highest) to 5 (Lowest) based on the existing evidence of its association with lifespan and longevity (see Supplementary Section SI.I for classification protocol). While only 26 genes have the highest confidence level, indicating that changes in their activity extend mammalian lifespan, most genes have low confidence and show a weak association with aging. \textbf{b,} The number of genes associated with each of the hallmarks of aging. 1,250 genes are associated by the OpenGenes database with one or more hallmarks of aging based on their biological role, while the remaining 1,108 genes are unclassified. Stars indicate the number of genes with a specific confidence level (1-5) associated with each hallmark. Note that the Exhaustion of stem cells and Changes in the extracellular matrix structure hallmarks are not associated with any of the genes in levels 1 and 2 while the Disabled macroautophagy hallmark is not associated with any of the genes in level 1. \textbf{c,}  The number of genes associated with multiple hallmarks. Reflecting the interconnectedness between the hallmarks, some of the genes are associated with multiple hallmarks. 1,108 genes are not linked to any hallmark, 860 genes are linked to a single hallmark, and 390 genes are shared by multiple hallmarks. The TP53 gene is associated with the most (7) hallmarks, reflecting its critical roles in various essential cellular processes such as DNA repair and apoptosis.}
\includegraphics[width=0.5\textwidth]{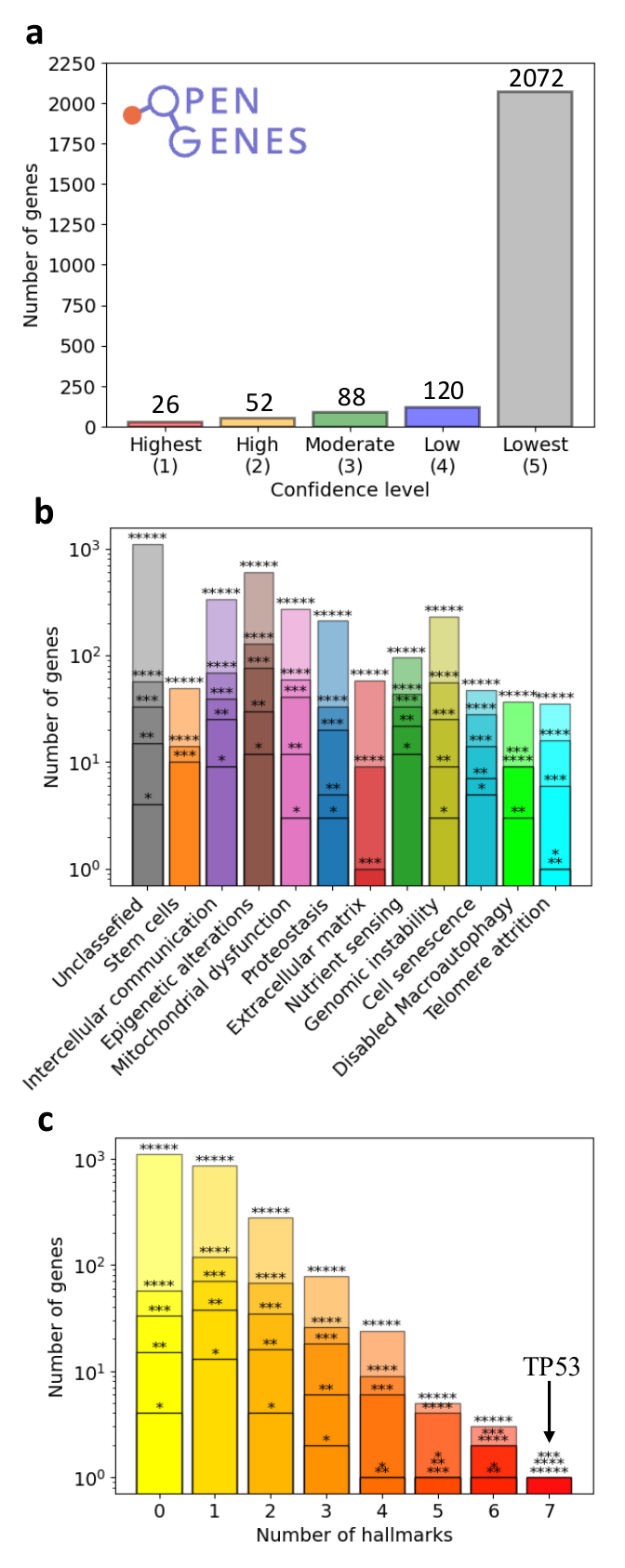}
\label{fig:Aging genes}
\end{SCfigure}

\clearpage

\begin{figure*}
	\centering
	\begin{tikzpicture}[      
	every node/.style={anchor=north east,inner sep=0pt},
	x=1mm, y=1mm,
	]   
	\node (fig1) at (0,0)
	{\includegraphics[scale=0.6]{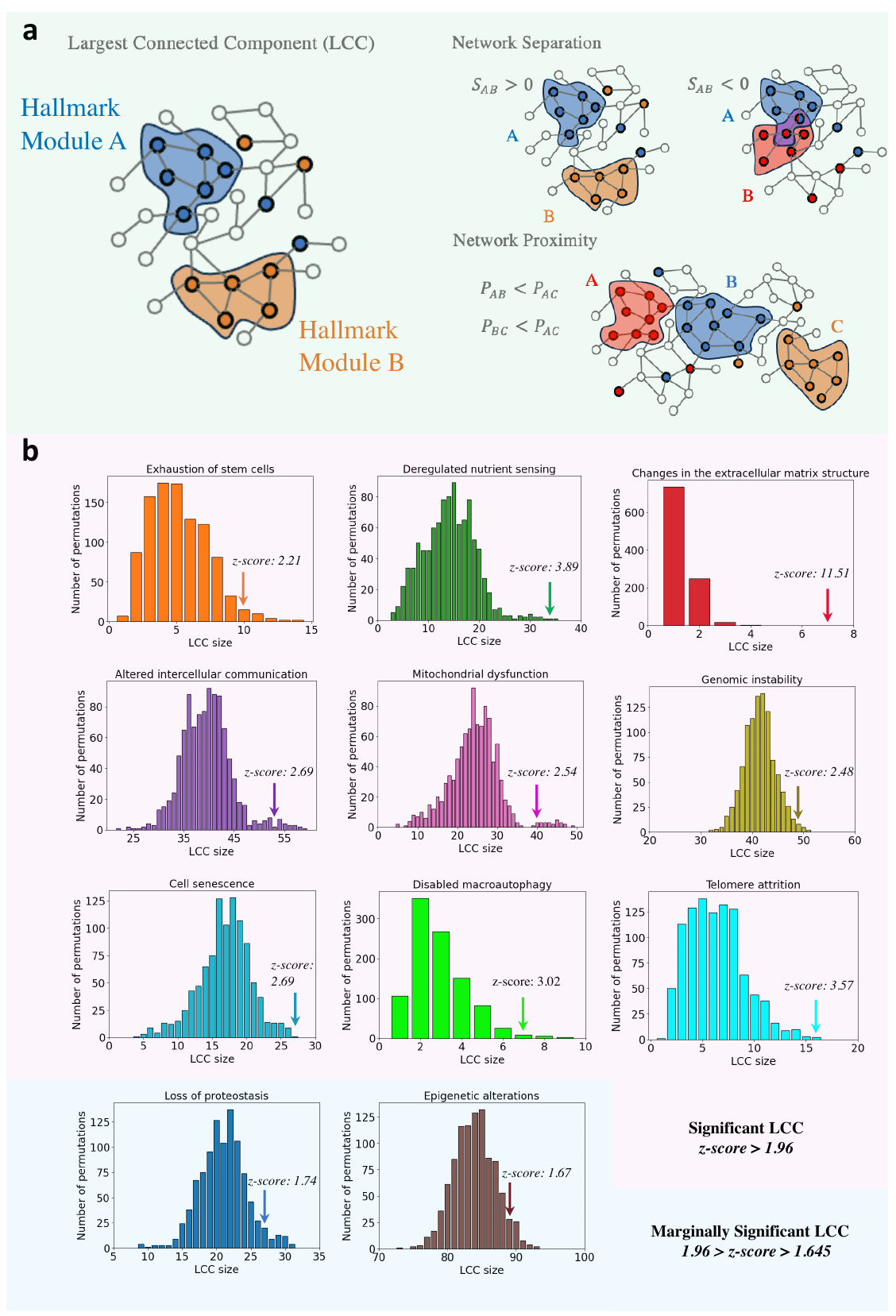}};
	\end{tikzpicture}
	\caption{\textbf{Network characteristics of the hallmarks of aging. a,} Genes associated with similar biological mechanisms often form connected components \cite{menche2015uncovering}. The largest connected component (LCC) characterizes the module of the gene set. The network structure allows us to analyze the network separation and proximity between different gene sets. Negative separation indicates overlapping modules while positive separation denotes distinct, topologically non-overlapping modules. Network proximity estimates the network-based distance between modules utilizing the shortest paths between pairs of genes in different modules. The proximity allows for estimating modules in close neighborhoods compared to distant ones. \textbf{b,} The LCC size and significance of each hallmark of aging compared to the distribution of LCCs formed by a randomized control group (see methods). The LCC formed by the genes of each hallmark defines the \textit{hallmark module}, characterizing the network representation of the hallmark. The genes of each hallmark of aging form a statistically significant LCC (defined as z-score $ > 1.96$) compared to the control group. The only two exceptions are the Loss of proteostasis hallmark (z-score $ = 1.74$) and the Epigenetic alterations hallmark (z-score $ = 1.67$) display marginal significance (defined as z-score $ > 1.645$).  }
	\label{fig:hallmarks_relations}	
\end{figure*}


\begin{figure}
	\centering
	\begin{tikzpicture}[      
	every node/.style={anchor=north east,inner sep=0pt},
	x=1mm, y=1mm,
	]   
	\node (fig1) at (0,0)
	{\includegraphics[scale=0.61]{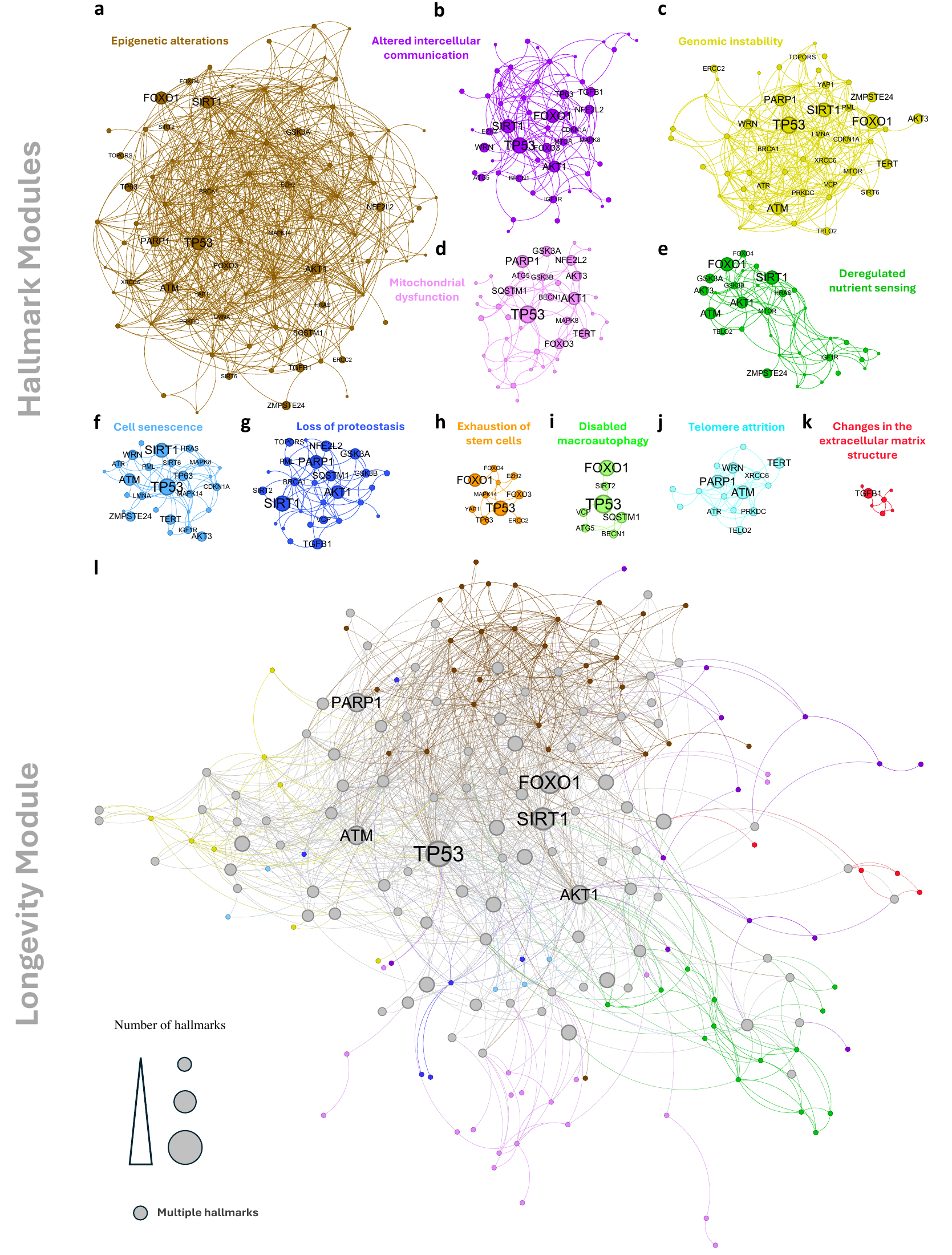}};
    \end{tikzpicture}
\end{figure}

\begin{figure}[t!]
    \caption{\textbf{The hallmark modules and the longevity module. a-k,} Genes associated with each of the hallmarks of aging are not randomly distributed in the human interactome but agglomerate in specific network neighborhood, forming a statistically significant LCC. These LCCs are the \textit{hallmark modules} -  sub-graphs of the human interactome representing the biological origin of the hallmarks of aging. Each hallmark module is shown separately in the figure with a distinct color. Genes associated with more than a single hallmark are shown with a label. \textbf{a,} Epigenetic alterations (89 genes). \textbf{b,} Altered intercellular communication (53 genes). \textbf{c,} Genomic instability (49 genes). \textbf{d,} Mitochondrial dysfunction (40 genes). \textbf{e,} Deregulated nutrient sensing (34 genes). \textbf{f,} Cell senescence (27 genes). \textbf{g,} Loss of proteostasis (27 genes). \textbf{h,} Exhaustion of stem cells (10 genes). \textbf{i,} Disabled macroautophagy (7 genes). \textbf{j,} Telomere attrition (16 genes). \textbf{k,} Changes in the extracellular matrix structure (7 genes). \textbf{l,} The 11 hallmark modules were found to be in the same network neighborhood and when agglomerated together form the \textit{longevity module} shown here. Genes associated with a single hallmark module are colored accordingly while genes associated with multiple hallmarks are colored in black with the node's edges colored based on their hallmark associations. The size of each node reflects its number of hallmark associations. The labels of genes associated with five or more hallmarks are shown.}
    \label{fig:longevity_module}
\end{figure}
\clearpage

\begin{figure*}
	\centering
	\begin{tikzpicture}[      
	every node/.style={anchor=north east,inner sep=0pt},
	x=1mm, y=1mm,
	]   
	\node (fig1) at (0,0)
	{\includegraphics[scale=0.43]{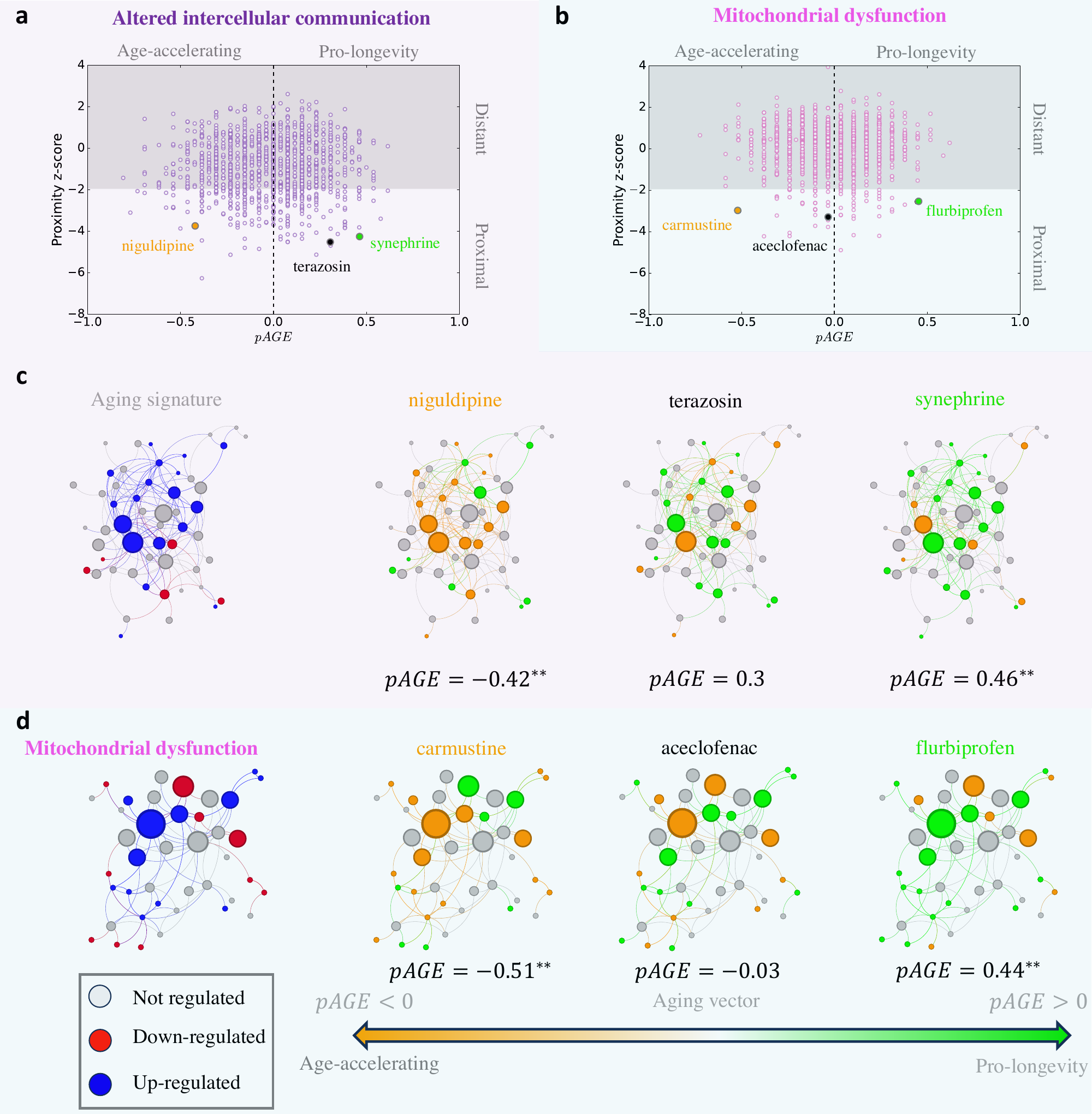}};
	\end{tikzpicture}
	\caption{\textbf{Perturbation pro-Age signature. a,b,} The network proximity identifies drugs with close targets to a specific hallmark module (z-score $<-1.96$), indicating a strong impact while the $pAGE$ values predict the direction of the impact to be beneficial ($pAGE > 0$) or deficient ($pAGE < 0$) for aging. The data points are all 1346 drugs that appear in both the DrugBank and CMap databases and their respective proximity z-score and $pAGE$ values for the Altered intercellular communication and Mitochondrial dysfunction hallmarks confidence level 4. The grey area indicates a lack of significance for proximity (z-score $> -1.96$). \textbf{c,d,} Comparing the aging signature to the drug signature allows for calculating the $pAGE$ value (Eq.~\eqref{eq:pAGE}) of drugs to each of the hallmarks of aging, predicting drugs that are beneficial ($pAGE > 0$, green) or deficient ($pAGE < 0$, orange) for aging and their respective significance.}
	\label{fig:signature_impact_directionality}	
\end{figure*}

\FloatBarrier

\begin{table}
	\centering
	\includegraphics[scale=1]{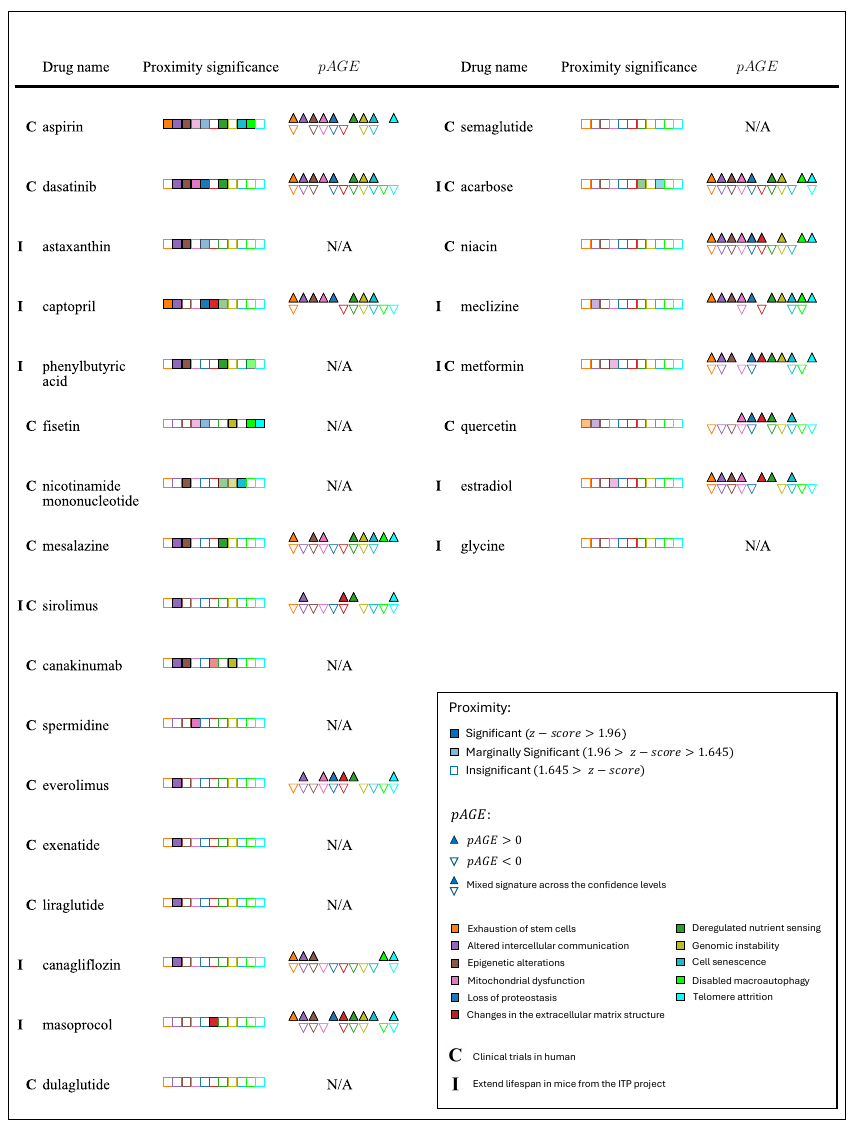}
    \caption{\textbf{Drug-repurposing of drugs currently under clinical trials for anti-aging and from the ITP project.} 25 drugs currently under clinical trials for anti-aging medicine for humans (\textbf{C}) \cite{guarente2023human} or found to extend lifespan in mice from the ITP project (\textbf{I}) \cite{miller2007aging}. Among them, 16 showed statistically significant proximity (z-score $<-1.96$) for at least one hallmark. Additional five shows marginal significance (z-score $<-1.645$). All drugs show positive $pAGE$ for at least four hallmarks. Statistically significant proximity is shown with full color and marginal significance proximity with transparent color. Non-significant are shown in white. Arrows indicate the $pAGE$ directionality (positive - up or negative - down). Proximity and $pAGE$ are measured across all confidence levels and the most significant result is shown.}
    \label{table:clinical_ITP_drug_repurposing}	
\end{table}

\begin{table}
	\centering
	\begin{tikzpicture}[      
	every node/.style={anchor=north east,inner sep=0pt},
	x=1mm, y=1mm,
	]   
	\node (fig1) at (0,0)
	{\includegraphics[scale=0.6]{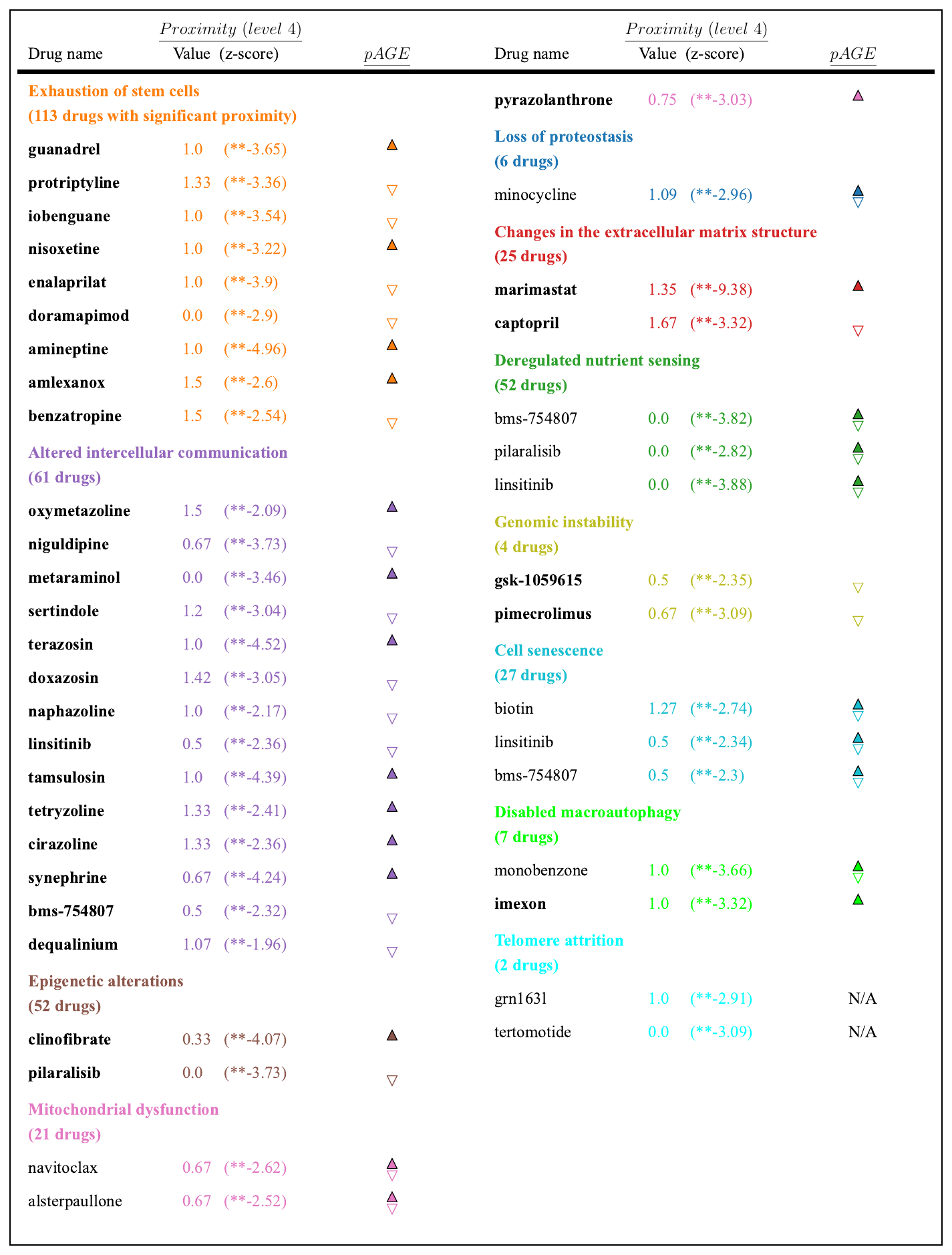}};
	\end{tikzpicture}
	\caption{\textbf{Drug-repurposing for Hallmark-targeted drugs.} For each hallmark, top candidates (partial list) with statistically significant proximity to each of the hallmarks of aging are shown. The proximity value and z-score are shown for level 4. Positive $pAGE$ values are shown as up and negative values as down. Mixed $pAGE$ signature across the confidence levels shown as both up and down.}
\label{table:top_drug_repurposing_each_hallmark}	
\end{table}

\begin{figure*}
	\centering
	\begin{tikzpicture}[      
	every node/.style={anchor=north east,inner sep=0pt},
	x=1mm, y=1mm,
	]   
	\node (fig1) at (0,0)
	{\includegraphics[scale=0.54]{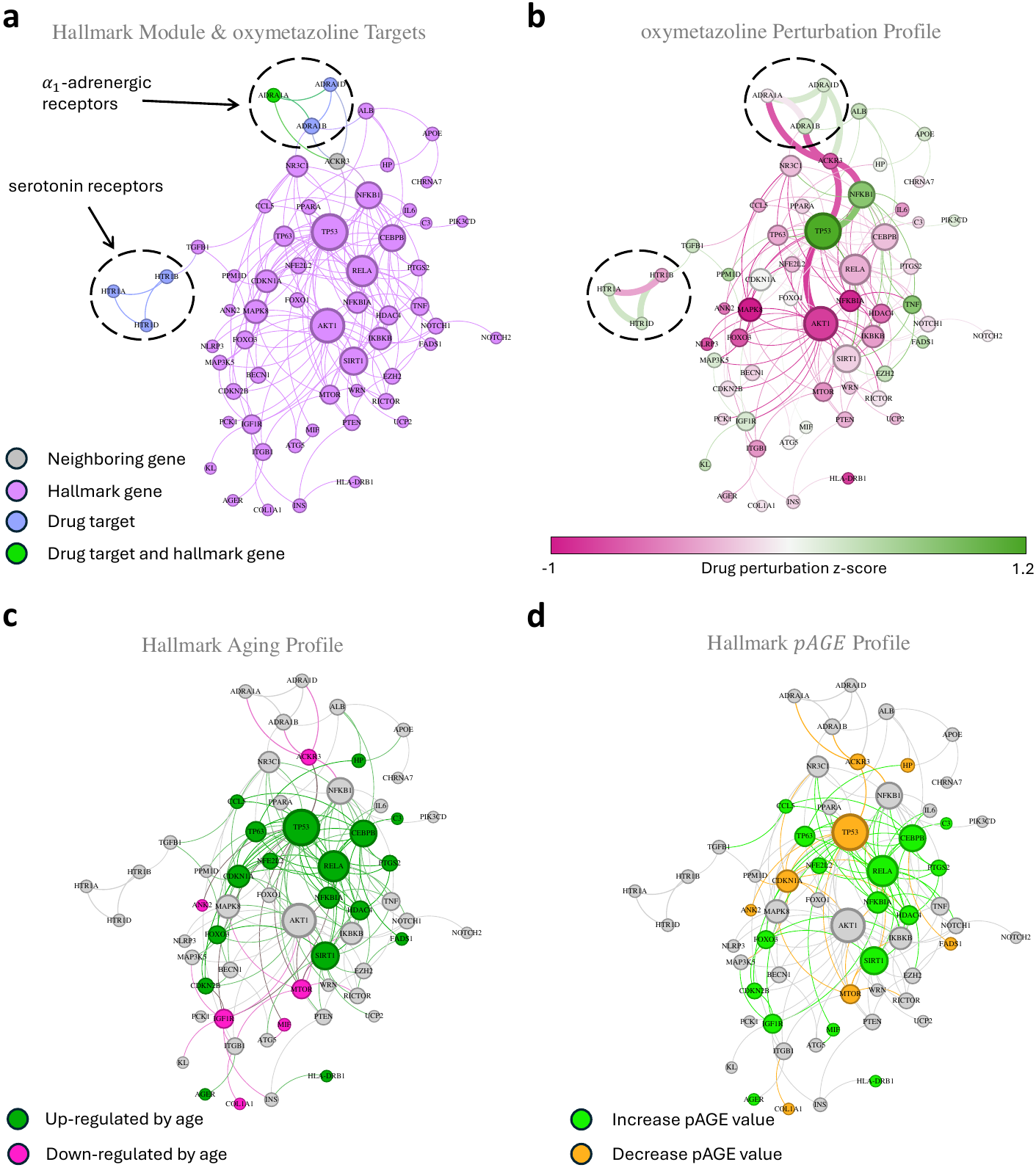}};
	\end{tikzpicture}
	\caption{\textbf{Proximity and $pAGE$ predicts mechanism of action. a,} Oxymetazoline is one of the candidates for drug-repurposing for the Altered intercellular communication hallmark (shown in purple). Oxymetazoline targets the proteins ADRA1A, ADRA1B, and ADRA1D as part of the $\alpha_1$-adrenergic receptor protein group, and the proteins HTR1A, HTR1B and HTR1D as part of the serotonin receptor protein group. The $\alpha_1$-adrenergic receptor protein group directly connects to the module by the ALB and NR3C1 genes, while the serotonin receptor protein group directly connects to the module by the TGFB1 gene. The gene ADRA1A is both a drug target and a hallmark gene (green) while ADRA1B and HTR1B (blue) are nearest neighbors of the module, leading to a statistically significant proximity of 1.5 (see Methods). \textbf{b,} Perturbing the MCF7 cell line with oxymetazoline, the drug signature up- (green) and down- (red) regulates genes in the module. The color bar shows the z-score of the perturbation signature for each gene. The perturbation follows a detour path starting from the $\alpha_1$-adrenergic receptor protein group, and does not follow the shortest path to the module through the immediate target neighbors ALB, and NR3C1. Instead, the target's neighbor ACKR3 (not a hallmark gene) is perturbed and transmits the information to the module. \textbf{c,} The aging signature marks genes that are up-regulated (green) and down-regulated (red) with age. \textbf{d,} The $pAGE$ value is measured according to Eq.~\eqref{eq:pAGE}. By comparing the aging signature and the drug signature, genes with opposite signs (green) increase the $pAGE$ value while genes with similar signs (orange) decrease it, resulting in a statistically significant $pAGE = 0.46$.}
	\label{fig:MOA}	
\end{figure*}



\FloatBarrier



\newpage


\printbibliography

@article{miller2007aging,
  title={An aging interventions testing program: study design and interim report},
  author={Miller, Richard A and Harrison, David E and Astle, Clinton M and Floyd, Robert A and Flurkey, Kevin and Hensley, Kenneth L and Javors, Martin A and Leeuwenburgh, Christiaan and Nelson, James F and Ongini, Ennio and others},
  journal={Aging cell},
  volume={6},
  number={4},
  pages={565--575},
  year={2007},
  publisher={Wiley Online Library}
}

@article{fu2021srt1720,
  title={SRT1720 Protects Against CSE-Induced Cellular Senescence via Accelerates of FOXO3-PINK1-mediated Mitophagy},
  author={Fu, Changyong and Yuan, Dong and Jiang, Yaona and Li, Yaqing and others},
  year={2021}
}

@article{hata2017synergistic,
  title={Synergistic activity and heterogeneous acquired resistance of combined MDM2 and MEK inhibition in KRAS mutant cancers},
  author={Hata, Aaron N and Rowley, Steve and Archibald, Hannah L and Gomez-Caraballo, Maria and Siddiqui, Faria M and Ji, Fei and Jung, Joonil and Light, Madelyn and Lee, Joon Sang and Debussche, Laurent and others},
  journal={Oncogene},
  volume={36},
  number={47},
  pages={6581--6591},
  year={2017},
  publisher={Nature Publishing Group}
}

@inproceedings{kersting2024nextflow,
  title={A Nextflow Pipeline for Network-Based Disease Module Identification and Validation},
  author={Kersting, Johannes and Manz, Quirin and Aguirre-Plans, Joaquim and Bucheron, Chlo{\'e} and Spindler, Lisa and Pock, Tanja and Delgado-Chaves, Fernando Miguel and Guney, Emre and List, Markus},
  booktitle={RExPO24 Conference},
  year={2024},
  organization={REPO4EU}
}

@article{spector2025transformers,
  title={Transformers Enhance the Predictive Power of Network Medicine},
  author={Spector, Jonah and Aldana, Andr{\'e}s and Sebek, Michael and Ehlert, Joey and De Frondeville, Christian and Ghiassian, Susan Dina and Barabasi, Albert-Laszlo},
  journal={medRxiv},
  pages={2025--01},
  year={2025},
  publisher={Cold Spring Harbor Laboratory Press}
}

@article{do2021network,
  title={Network medicine framework shows that proximity of polyphenol targets and disease proteins predicts therapeutic effects of polyphenols},
  author={Do Valle, Italo F and Roweth, Harvey G and Malloy, Michael W and Moco, Sofia and Barron, Denis and Battinelli, Elisabeth and Loscalzo, Joseph and Barabasi, Albert-Laszlo},
  journal={Nature Food},
  volume={2},
  number={3},
  pages={143--155},
  year={2021},
  publisher={Nature Publishing Group UK London}
}

@article{subramanian2017next,
  title={A next generation connectivity map: L1000 platform and the first 1,000,000 profiles},
  author={Subramanian, Aravind and Narayan, Rajiv and Corsello, Steven M and Peck, David D and Natoli, Ted E and Lu, Xiaodong and Gould, Joshua and Davis, John F and Tubelli, Andrew A and Asiedu, Jacob K and others},
  journal={Cell},
  volume={171},
  number={6},
  pages={1437--1452},
  year={2017},
  publisher={Elsevier}
}

@article{barabasi2011network,
  title={Network medicine: a network-based approach to human disease},
  author={Barab{\'a}si, Albert-L{\'a}szl{\'o} and Gulbahce, Natali and Loscalzo, Joseph},
  journal={Nature Reviews Genetics},
  volume={12},
  number={1},
  pages={56--68},
  year={2011},
  publisher={Nature Publishing Group}
}

@article{menche2015uncovering,
  title={Uncovering disease-disease relationships through the incomplete interactome},
  author={Menche, J{\"o}rg and others},
  journal={Science},
  volume={347},
  number={6224},
  pages={1257601},
  year={2015},
  publisher={American Association for the Advancement of Science}
}

@article{patel2017oxymetazoline,
  title={Oxymetazoline hydrochloride cream for facial erythema associated with rosacea},
  author={Patel, Nupur U and Shukla, Shweta and Zaki, Jessica and Feldman, Steven R},
  journal={Expert review of clinical pharmacology},
  volume={10},
  number={10},
  pages={1049--1054},
  year={2017},
  publisher={Taylor \& Francis}
}

@article{lopez2023hallmarks,
  title={Hallmarks of aging: An expanding universe},
  author={L{\'o}pez-Ot{\'\i}n, Carlos and Blasco, Maria A and Partridge, Linda and Serrano, Manuel and Kroemer, Guido},
  journal={Cell},
  year={2023},
  publisher={Elsevier}
}

@article{patten2022identification,
  title={Identification of potent inhibitors of SARS-CoV-2 infection by combined pharmacological evaluation and cellular network prioritization},
  author={Patten, JJ and others},
  journal={Iscience},
  volume={25},
  number={9},
  year={2022},
  publisher={Elsevier}
}

@article{doze2011long,
  title={Long-term $\alpha$1A-adrenergic receptor stimulation improves synaptic plasticity, cognitive function, mood, and longevity},
  author={Doze, Van A and Papay, Robert S and Goldenstein, Brianna L and Gupta, Manveen K and Collette, Katie M and Nelson, Brian W and Lyons, Mariaha J and Davis, Bethany A and Luger, Elizabeth J and Wood, Sarah G and others},
  journal={Molecular pharmacology},
  volume={80},
  number={4},
  pages={747--758},
  year={2011},
  publisher={Elsevier}
}

@article{lagunas2022g,
  title={G protein-coupled receptors that influence lifespan of human and animal models},
  author={Lagunas-Rangel, Francisco Alejandro},
  journal={Biogerontology},
  volume={23},
  number={1},
  pages={1--19},
  year={2022},
  publisher={Springer}
}

@article{goeminne2024plasma,
  title={Plasma protein-based organ-specific aging and mortality models unveil diseases as accelerated aging of organismal systems},
  author={Goeminne, Ludger JE and Vladimirova, Anastasiya and Eames, Alec and Tyshkovskiy, Alexander and Argentieri, M Austin and Ying, Kejun and Moqri, Mahdi and Gladyshev, Vadim N},
  journal={Cell Metabolism},
  year={2024},
  publisher={Elsevier}
}

@book{alon2023systems,
  title={Systems medicine: physiological circuits and the dynamics of disease},
  author={Alon, Uri},
  year={2023},
  publisher={CRC Press}
}

@article{mitchell2015animal,
  title={Animal models of aging research: implications for human aging and age-related diseases},
  author={Mitchell, Sarah J and Scheibye-Knudsen, Morten and Longo, Dan L and de Cabo, Rafael},
  journal={Annu. Rev. Anim. Biosci.},
  volume={3},
  number={1},
  pages={283--303},
  year={2015},
  publisher={Annual Reviews}
}

@article{guarente2000genetic,
  title={Genetic pathways that regulate ageing in model organisms},
  author={Guarente, Leonard and Kenyon, Cynthia},
  journal={Nature},
  volume={408},
  number={6809},
  pages={255--262},
  year={2000},
  publisher={Nature Publishing Group UK London}
}

@article{fong2024principal,
  title={Principal component-based clinical aging clocks identify signatures of healthy aging and targets for clinical intervention},
  author={Fong, Sheng and Pabis, Kamil and Latumalea, Djakim and Dugersuren, Nomuundari and Unfried, Maximilian and Tolwinski, Nicholas and Kennedy, Brian and Gruber, Jan},
  journal={Nature Aging},
  volume={4},
  number={8},
  pages={1137--1152},
  year={2024},
  publisher={Nature Publishing Group US New York}
}

@article{cohen2022complex,
  title={A complex systems approach to aging biology},
  author={Cohen, Alan A and Ferrucci, Luigi and F{\"u}l{\"o}p, Tam{\`a}s and Gravel, Dominique and Hao, Nan and Kriete, Andres and Levine, Morgan E and Lipsitz, Lewis A and Olde Rikkert, Marcel GM and Rutenberg, Andrew and others},
  journal={Nature aging},
  volume={2},
  number={7},
  pages={580--591},
  year={2022},
  publisher={Nature Publishing Group US New York}
}

@article{han2008understanding,
  title={Understanding biological functions through molecular networks},
  author={Han, Jing-Dong Jackie},
  journal={Cell research},
  volume={18},
  number={2},
  pages={224--237},
  year={2008},
  publisher={Nature Publishing Group}
}

@article{cheng2019network,
  title={Network-based prediction of drug combinations},
  author={Cheng, Feixiong and Kov{\'a}cs, Istv{\'a}n A and Barab{\'a}si, Albert-L{\'a}szl{\'o}},
  journal={Nature communications},
  volume={10},
  number={1},
  pages={1197},
  year={2019},
  publisher={Nature Publishing Group UK London}
}

@article{sharma2015disease,
  title={A disease module in the interactome explains disease heterogeneity, drug response and captures novel pathways and genes in asthma},
  author={Sharma, Amitabh and others},
  journal={Human molecular genetics},
  volume={24},
  number={11},
  pages={3005--3020},
  year={2015},
  publisher={Oxford University Press}
}

@article{finch2001genetics,
  title={The genetics of aging},
  author={Finch, Caleb E and Ruvkun, Gary},
  journal={Annual review of genomics and human genetics},
  volume={2},
  number={1},
  pages={435--462},
  year={2001},
  publisher={Annual Reviews 4139 El Camino Way, PO Box 10139, Palo Alto, CA 94303-0139, USA}
}

@article{finch1997genetics,
  title={Genetics of aging},
  author={Finch, Caleb E and Tanzi, Rudolph E},
  journal={Science},
  volume={278},
  number={5337},
  pages={407--411},
  year={1997},
  publisher={American Association for the Advancement of Science}
}

@article{brooks2013genetics,
  title={Genetics of healthy aging and longevity},
  author={Brooks-Wilson, Angela R},
  journal={Human genetics},
  volume={132},
  pages={1323--1338},
  year={2013},
  publisher={Springer}
}

@article{de2009meta,
  title={Meta-analysis of age-related gene expression profiles identifies common signatures of aging},
  author={De Magalh{\~a}es, Jo{\~a}o Pedro and Curado, Jo{\~a}o and Church, George M},
  journal={Bioinformatics},
  volume={25},
  number={7},
  pages={875--881},
  year={2009},
  publisher={Oxford University Press}
}

@article{pal2016epigenetics,
  title={Epigenetics and aging},
  author={Pal, Sangita and Tyler, Jessica K},
  journal={Science advances},
  volume={2},
  number={7},
  pages={e1600584},
  year={2016},
  publisher={American Association for the Advancement of Science}
}

@article{newman2010meta,
  title={A meta-analysis of four genome-wide association studies of survival to age 90 years or older: the Cohorts for Heart and Aging Research in Genomic Epidemiology Consortium},
  author={Newman, Anne B and others},
  journal={Journals of Gerontology Series A: Biomedical Sciences and Medical Sciences},
  volume={65},
  number={5},
  pages={478--487},
  year={2010},
  publisher={Oxford University Press}
}

@article{guthrie2023autocore,
  title={AutoCore: A network-based definition of the core module of human autoimmunity and autoinflammation},
  author={Guthrie, Julia and K{\"o}stel Bal, Sevgi and Lombardo, Salvo Danilo and M{\"u}ller, Felix and Sin, Celine and H{\"u}tter, Christiane VR and Menche, J{\"o}rg and Boztug, Kaan},
  journal={Science Advances},
  volume={9},
  number={35},
  pages={eadg6375},
  year={2023},
  publisher={American Association for the Advancement of Science}
}

@article{zenin2019identification,
  title={Identification of 12 genetic loci associated with human healthspan},
  author={Zenin, Aleksandr and others},
  journal={Communications biology},
  volume={2},
  number={1},
  pages={1--11},
  year={2019},
  publisher={Nature Publishing Group}
}

@article{wang2022epigenetic,
  title={Epigenetic regulation of aging: implications for interventions of aging and diseases},
  author={Wang, Kang and others},
  journal={Signal transduction and targeted therapy},
  volume={7},
  number={1},
  pages={374},
  year={2022},
  publisher={Nature Publishing Group UK London}
}

@article{barzilai2012critical,
  title={The critical role of metabolic pathways in aging},
  author={Barzilai, Nir and Huffman, Derek M and Muzumdar, Radhika H and Bartke, Andrzej},
  journal={Diabetes},
  volume={61},
  number={6},
  pages={1315--1322},
  year={2012},
  publisher={Am Diabetes Assoc}
}

@article{kanehisa2000kegg,
  title={KEGG: kyoto encyclopedia of genes and genomes},
  author={Kanehisa, Minoru and Goto, Susumu},
  journal={Nucleic acids research},
  volume={28},
  number={1},
  pages={27--30},
  year={2000},
  publisher={Oxford University Press}
}

@article{peters2015transcriptional,
  title={The transcriptional landscape of age in human peripheral blood},
  author={Peters, Marjolein J and others},
  journal={Nature communications},
  volume={6},
  number={1},
  pages={1--14},
  year={2015},
  publisher={Nature Publishing Group}
}

@article{jia2018analysis,
  title={An analysis of aging-related genes derived from the genotype-tissue expression project (GTEx)},
  author={Jia, Kaiwen and others},
  journal={Cell death discovery},
  volume={4},
  number={1},
  pages={91},
  year={2018},
  publisher={Nature Publishing Group UK London}
}

@article{lonsdale2013genotype,
  title={The genotype-tissue expression (GTEx) project},
  author={Lonsdale, John and Thomas, Jeffrey and Salvatore, Mike and Phillips, Rebecca and Lo, Edmund and Shad, Saboor and Hasz, Richard and Walters, Gary and Garcia, Fernando and Young, Nancy and others},
  journal={Nature genetics},
  volume={45},
  number={6},
  pages={580--585},
  year={2013},
  publisher={Nature Publishing Group}
}

@article{timmers2020multivariate,
  title={Multivariate genomic scan implicates novel loci and haem metabolism in human ageing},
  author={Timmers, Paul RHJ and Wilson, James F and Joshi, Peter K and Deelen, Joris},
  journal={Nature communications},
  volume={11},
  number={1},
  pages={3570},
  year={2020},
  publisher={Nature Publishing Group UK London}
}

@article{timmers2019genomics,
  title={Genomics of 1 million parent lifespans implicates novel pathways and common diseases and distinguishes survival chances},
  author={Timmers, Paul RHJ and others},
  journal={elife},
  volume={8},
  pages={e39856},
  year={2019},
  publisher={eLife Sciences Publications, Ltd}
}

@article{guarente2023human,
  title={Human trials exploring anti-aging medicines},
  author={Guarente, Leonard and Sinclair, David A and Kroemer, Guido},
  journal={Cell Metabolism},
  year={2023},
  publisher={Elsevier}
}

@article{wishart2006drugbank,
  title={DrugBank: a comprehensive resource for in silico drug discovery and exploration},
  author={Wishart, David S and Knox, Craig and Guo, An Chi and Shrivastava, Savita and Hassanali, Murtaza and Stothard, Paul and Chang, Zhan and Woolsey, Jennifer},
  journal={Nucleic acids research},
  volume={34},
  number={suppl\_1},
  pages={D668--D672},
  year={2006},
  publisher={Oxford University Press}
}

@article{morselli2021network,
  title={Network medicine framework for identifying drug-repurposing opportunities for COVID-19},
  author={Morselli Gysi, Deisy and Do Valle, {\'I}talo and Zitnik, Marinka and Ameli, Asher and Gan, Xiao and Varol, Onur and Ghiassian, Susan Dina and Patten, JJ and Davey, Robert A and Loscalzo, Joseph and others},
  journal={Proceedings of the National Academy of Sciences},
  volume={118},
  number={19},
  pages={e2025581118},
  year={2021},
  publisher={National Acad Sciences}
}

@article{guney2016network,
  title={Network-based in silico drug efficacy screening},
  author={Guney, Emre and Menche, J{\"o}rg and Vidal, Marc and Bar{\'a}basi, Albert-L{\'a}szl{\'o}},
  journal={Nature communications},
  volume={7},
  number={1},
  pages={10331},
  year={2016},
  publisher={Nature Publishing Group UK London}
}

@article{burla2018genomic,
  title={Genomic instability and DNA replication defects in progeroid syndromes},
  author={Burla, Romina and La Torre, Mattia and Merigliano, Chiara and Verni, Fiammetta and Saggio, Isabella},
  journal={Nucleus},
  volume={9},
  number={1},
  pages={368--379},
  year={2018},
  publisher={Taylor \& Francis}
}

@article{kipling2004can,
  title={What can progeroid syndromes tell us about human aging?},
  author={Kipling, David and Davis, Terence and Ostler, Elizabeth L and Faragher, Richard GA},
  journal={Science},
  volume={305},
  number={5689},
  pages={1426--1431},
  year={2004},
  publisher={American Association for the Advancement of Science}
}

@article{freitas2011data,
  title={A data mining approach for classifying DNA repair genes into ageing-related or non-ageing-related},
  author={Freitas, Alex A and Vasieva, Olga and de Magalh{\~a}es, Jo{\~a}o Pedro},
  journal={BMC genomics},
  volume={12},
  pages={1--11},
  year={2011},
  publisher={Springer}
}

@article{schumacher2008age,
  title={Age to survive: DNA damage and aging},
  author={Schumacher, Bj{\"o}rn and Garinis, George A and Hoeijmakers, Jan HJ},
  journal={Trends in Genetics},
  volume={24},
  number={2},
  pages={77--85},
  year={2008},
  publisher={Elsevier}
}

@article{gensler1981dna,
  title={DNA damage as the primary cause of aging},
  author={Gensler, Helen L and Bernstein, Harris},
  journal={The Quarterly review of biology},
  volume={56},
  number={3},
  pages={279--303},
  year={1981},
  publisher={Stony Brook Foundation, Inc.}
}

@article{sebastiani2021protein,
  title={Protein signatures of centenarians and their offspring suggest centenarians age slower than other humans},
  author={Sebastiani, Paola and others},
  journal={Aging cell},
  volume={20},
  number={2},
  pages={e13290},
  year={2021},
  publisher={Wiley Online Library}
}

@article{aging2021aging,
  title={Aging Atlas: a multi-omics database for aging biology},
  journal={Nucleic acids research},
  volume={49},
  number={D1},
  pages={D825--D830},
  year={2021},
  publisher={Oxford University Press}
}

@article{gems2021hoverfly,
  title={The hoverfly and the wasp: A critique of the hallmarks of aging as a paradigm},
  author={Gems, David and De Magalh{\~a}es, Jo{\~a}o Pedro},
  journal={Ageing research reviews},
  volume={70},
  pages={101407},
  year={2021},
  publisher={Elsevier}
}

@article{jaccard1901etude,
  title={{\'E}tude comparative de la distribution florale dans une portion des Alpes et des Jura},
  author={Jaccard, Paul},
  journal={Bull Soc Vaudoise Sci Nat},
  volume={37},
  pages={547--579},
  year={1901}
}

@article{pun2022hallmarks,
  title={Hallmarks of aging-based dual-purpose disease and age-associated targets predicted using PandaOmics AI-powered discovery engine},
  author={Pun, Frank W and others},
  journal={Aging (Albany NY)},
  volume={14},
  number={6},
  pages={2475},
  year={2022},
  publisher={Impact Journals, LLC}
}

@article{lopez2013hallmarks,
  title={The hallmarks of aging},
  author={L{\'o}pez-Ot{\'\i}n, Carlos and Blasco, Maria A and Partridge, Linda and Serrano, Manuel and Kroemer, Guido},
  journal={Cell},
  volume={153},
  number={6},
  pages={1194--1217},
  year={2013},
  publisher={Elsevier}
}

@article{rafikova2023open,
  title={Open Genes—a new comprehensive database of human genes associated with aging and longevity},
  author={Rafikova, Ekaterina and Nemirovich-Danchenko, Nikolay and Ogmen, Anna and Parfenenkova, Anna and Velikanova, Anastasiia and Tikhonov, Stanislav and Peshkin, Leonid and Rafikov, Konstantin and Spiridonova, Olga and Belova, Yulia and others},
  journal={Nucleic Acids Research},
  volume={52},
  number={D1},
  pages={D950--D962},
  year={2024},
  publisher={Oxford University Press}
}
\end{document}